\documentclass[aps,pra,twocolumn,floats]{revtex4}

\usepackage{graphicx}
\usepackage{amsmath}
\usepackage{amssymb}
\usepackage[usenames,dvipsnames]{color}
\usepackage{textcomp}
\usepackage[normalem]{ulem}
\usepackage{bm}
\newcommand{\E}{{\cal E}}

\newcommand{\D}{{\Delta}}

\newcommand {\ket}[1]{|\,{#1}\,\rangle}

\newcommand{\beq}{\begin{equation}}
\newcommand{\eeq}{\end{equation}}
\newcommand{\bea}{\begin{eqnarray}}
\newcommand{\eea}{\end{eqnarray}}

\begin{document}
\title{Experimental observation of Anderson localization in laser-kicked  molecular rotors.}

\date{\today}
\author{M.~Bitter and V.~Milner}
\affiliation{Department of  Physics \& Astronomy and The
Laboratory for Advanced Spectroscopy and Imaging Research
(LASIR), The University of British Columbia, Vancouver, Canada \\}

%------------------------Abstract---------------------------------------------
\begin{abstract}{

We observe and study the phenomenon of Anderson localization in a system of true quantum kicked rotors. Nitrogen molecules in a supersonic molecular jet are cooled down to  27~K and are rotationally excited by a periodic train of 24~high-intensity femtosecond pulses. Exponential distribution of the molecular angular momentum - the most unambiguous signature of Anderson localization - is measured directly by means of coherent Raman scattering. We demonstrate the suppressed growth of the molecular rotational energy with the number of laser kicks and study the dependence of the localization length on the kick strength. Both timing and amplitude noise in the pulse train is shown to destroy the localization and revive the diffusive growth of angular momentum.}
\end{abstract}

\maketitle

%====================================================================
%                    INTRODUCTION
%====================================================================

%-------------Intro 1: Anderson localization in QKR -----------------

The periodically kicked rotor is one of the simplest systems whose classical motion is chaotic, as manifested by the unbounded diffusive growth of its energy with the number of kicks. In contrast, the energy growth of a quantum kicked rotor (QKR) is suppressed due to the interference of quantum interaction pathways \cite{Casati1979, Izrailev1980}. The effect has been linked to Anderson localization \cite{Anderson1958} of the electronic wave function in disordered solids \cite{Fishman1982}. Similarly to the latter, the wave function of the quantum rotor does not grow wider in the angular momentum space with every consecutive kick, but instead localizes near the initial rotational state, with the probability amplitude falling exponentially away from it.

%-------------Intro 2: Previous Work on QKR -------------------------

The exponential distribution around the localization center is considered a necessary component and a distinct signature of Anderson localization. Although it has been demonstrated \cite{Moore1995} in a cold-atom analogue of the QKR \cite{Raizen1999}, exponentially localized states have not yet been observed in a system of true quantum rotors. A natural choice for such a system - a diatomic molecule subject to short kicks from a pulsed external field (microwave, optical or THz), has been discussed in multiple theoretical proposals \cite{Bluemel1986, Gong2001, Floss2012, Matsuoka2015}. In a series of recent works \cite{Floss2012, Floss2013, Floss2014, Floss2015b}, Averbukh and coworkers suggested a strategy to observe and study a number of QKR effects in an ensemble of molecules exposed to a periodic sequence of ultra-short laser pulses. The effects of a quantum resonance \cite{Cryan2009, Zhdanovich2012} and Bloch oscillations \cite{Floss2015a} have been verified experimentally. An onset of Anderson localization in laser-induced molecular alignment has been reported \cite{Kamalov2015}, but the direct evidence of the exponentially localized states and the suppressed growth of the rotational energy has not been shown.

%-------------Intro 3: Challenges of observing AL -------------------

The difficulty of demonstrating Anderson localization with molecular rotors stems from a number of experimental challenges. First, the need to assess the shape of the rotational distribution calls for a sensitive detection method capable of resolving individual rotational states. According to the theoretical studies \cite{Floss2013}, the population of a few tens of rotational states must be measured with high sensitivity over the range of at least two orders of magnitude. Second, for the localized state not to be smeared out due to the averaging over the initial thermal distribution, the latter must be narrowed down to as close to a single rotational state as possible, requiring cold molecular samples. Finally, an important test of Anderson localization, the recovery of classical diffusion under the influence of noise and decoherence, demonstrated experimentally with cold atoms \cite{Klappauf1998, Ammann1998, Milner2000, Oskay2003} and theoretically with molecular QKR \cite{Floss2013}, requires long sequences of more than 20 strong kicks.

%-------------Intro 3: Punch-line paragraph -------------------------

In this work, we address all three of the above challenges and study the rotational dynamics of nitrogen molecules, cooled down to 27~K in a supersonic expansion and kicked by a periodic series of 24 laser pulses. We use state-resolved coherent Raman spectroscopy to demonstrate the exponential shape of the created rotational wave packet, indicative of Anderson localization. The dependence of the rotational distribution on the number of pulses and their strength is investigated. Our ability to resolve individual rotational states allows for a direct extraction of the absorbed energy, whose growth is shown to cease completely after as few as three pulses. To confirm the coherent nature of the observed localization, we study the effect of both timing noise and amplitude noise, which are shown to yield a non-exponential distribution of angular momenta and revive the diffusive growth of energy. Our results are in good agreement with the theoretical analysis of Flo{\ss}, Fishman and Averbukh \cite{Floss2013} and our own numerical simulations.

%====================================================================
%                    SETUP
%====================================================================

\begin{figure}
\centering
 \includegraphics[width=1.0\columnwidth]{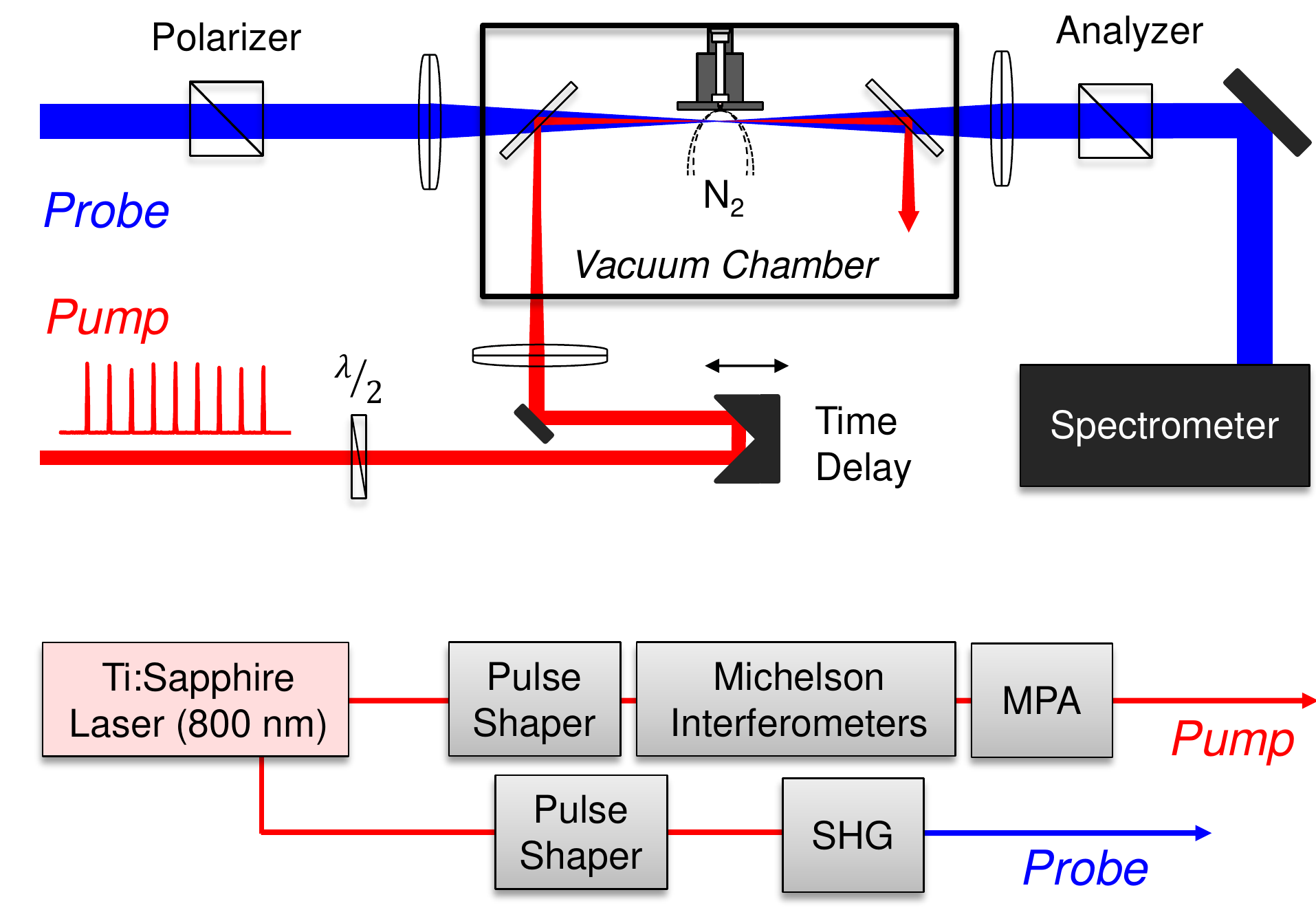}
     \caption{ (top) Scheme of the experimental setup. A train of strong femtosecond pulses (pump) and a delayed weak probe pulse are focused on a supersonic jet of nitrogen molecules in a vacuum chamber. The change of probe polarization is analyzed as a function of wavelength by means of two crossed polarizers and a spectrometer. (bottom) Diagram of the pump and probe sources. Long pulse trains are generated by the combination of a pulse shaper and two Michelson interferometers; their energy is then boosted by a multi-pass amplifier (MPA). Another pulse shaper is used to narrow the spectral bandwidth of probe pulses, whose central wavelength is shifted by means of second harmonic generation (SHG) in a nonlinear crystal.}
  \vskip -.1truein
  \label{Fig:Setup}
\end{figure}

%--------- Setup 1: High-intensity pulse train ----------------------

To expose nitrogen molecules to a long series of strong kicks, we have developed an optical setup capable of generating high-intensity trains of femtosecond pulses, shown schematically in the lower panel of Fig.\ref{Fig:Setup} and described in detail in \cite{Bitter2016a}. Briefly, we use a Ti:Sapphire femtosecond laser system producing pulses of 130~fs full width at half maximum (FWHM) at a central wavelength of 800~nm, 1~kHz repetition rate, and 2~mJ per pulse. Part of the beam ($40\%$ in energy) is sent through a standard `$4f$' pulse shaper \cite{Weiner2000}, which generates a sequence of six pulses separated by arbitrary time intervals in a total window of 50~ps. The shaper is followed by two Michelson interferometers quadrupling the amount of laser pulses. The entire sequence of 24~weak pulses is amplified by a home-built multi-pass amplifier (MPA) boosting the optical energy to more than $100~\mathrm{\mu J}$ per pulse at 10~Hz repetition rate. The standard deviation of the pulse energy fluctuations is below $15\%$.

%---------Setup 2: Supersonic jet -----------------------------------

The remaining part of the 800~nm beam is used as a probe. Its spectrum is narrowed down in a separate pulse shaper, before its central wavelength is shifted to $\approx 400$~nm by frequency doubling. The probe pulses of 0.15~nm spectral width (FWHM) are linearly polarized at 45\textdegree\  with respect to the pulses in the pump train. Both beams are focused into a vacuum chamber, where they are combined on a dichroic beamsplitter and intersect a supersonic jet of N$_2$, as illustrated in the upper panel of Fig.~\ref{Fig:Setup}. Special care is taken to avoid detrimental effects of spatial averaging by making the probe beam significantly smaller than the pump (FWHM beam diameters of $20~\mathrm{\mu m}$ and $60~\mathrm{\mu m}$, respectively). We use a $500~\mathrm{\mu m}$ diameter pulsed nozzle, operating at the repetition rate of 10~Hz and the stagnation pressure of 33~bar, to achieve the rotational temperature of $27\pm3$~K at a distance of 1~mm from the nozzle \cite{Temperature}.

%---------Setup 3: Raman Spectroscopy -------------------------------

Coherent molecular rotation, produced by a strong pulse train, modulates the refractive index of the gas. As a result, the spectrum of a weak probe light acquires Raman sidebands shifted with respect to its central frequency and polarized orthogonally to its initial polarization \cite{Korech2013, Korobenko2014a}. By passing the probe pulses through an analyzer set at 90\textdegree\  with respect to the input probe polarization, we detect the rotational Raman spectrum of kicked molecules with a dynamic range of four orders of magnitude in intensity.

Consider a coherent superposition of two rotational states, $\psi_{J,M}(t)=c_{J,M} \ e^{- 2\pi i E_J t/h} \  \ket{J,M} + c_{J+2,M} \ e^{-2\pi i E_{J+2} t/h} \  \ket{J+2,M}$, created by a linearly polarized pump field. Here $J$ and $M$ are the molecular angular momentum and its projection on the vector of pump polarization, $E_J=hcBJ(J+1)$ is the rotational energy of a rigid rotor with the rotational constant $B$, $c$ is the speed of light in vacuum and $h$ is the Planck's constant. The coherent dynamics of such a wave packet will result in a Raman peak with a $J$-dependent frequency shift $\Delta \omega_{J}=(E_{J+2}-E_J)/h=2Bc(2J+3)$. Owing to the selection rules for a two-photon excitation process, $\Delta J=0,\pm2 \text{ and } \Delta M=0$, the superposition $\psi_{J,M}(t)$ can originate from any initially populated thermal state $\ket{J'=J\pm 2 k,M'=M}$, where $k$ is an integer. Hence, the intensity of the observed Raman peak will be proportional to the modulus squared of the induced coherence, $I_J \propto \sum_M \langle |c^{*}_{J,M} c^{}_{J+2,M}|^2 \rangle_{_{J'M'}}$, summed over the degenerate $M$-sublevels and averaged over the initial thermal mixture.

Note that if the initial ensemble contained only one populated level $|J'=J_0,M'=M_0\rangle$, the strength of the Raman signal would reduce to $I_J \propto P_{J,M_0} P_{J+2,M_0}$, where $P_{J,M}= |c^{}_{J,M}|^2 $ is the rotational population. For localized and non-localized dynamics of the QKR, we expect exponential or Gaussian population distributions, respectively \cite{Cohen1991, Klappauf1998}.
In either case, the Raman spectrum can be further simplified to $I_J \propto (P_{J,M_0})^2$, offering the direct measure of the rotational population. As we show below, this proportionality holds even at a non-zero temperature of molecules in our supersonic jet, when the Raman signal is produced by a number of independent rotational wave packets originating from different initial states $\ket{J',M'}$.
At 27~K most population is initially at $J'=2$. Thus, the smallness of $M'=0,\pm1,\pm2$ with respect to the angular momentum of the majority of states in the final wave packet results in an interaction Hamiltonian which to a good degree of approximation does not depend on $M'$ \cite{approximation}. Having all molecules in the thermal ensemble respond to the laser field in an almost identical way enables us to extract rotational populations from the Raman signal as $P_J = a_J \sqrt{I_J}$, with the coefficients $a_J$ found from normalizing the total population to unity.

%---------Setup 4: Kick strength calibration ------------------------

To determine the exact pulse intensity in the interaction region, we tune the period of the pulse train to the rotational period of a wave packet consisting of two rotational states with $J=2$ and $J=4$ \cite{Bitter2016b}. Fitting the frequency of the ensuing Rabi oscillations between the two states provides an accurate way of measuring the intensity of the pump pulses \cite{Pcalibration}. It is often expressed in the dimensionless units of ``kick strength'' $P= \D \alpha /(4\hbar) \int \E^2(t) dt$, where $\D\alpha$ is the polarizability anisotropy of the molecule and $\E$ the temporal envelope of the pulse. The kick strength reflects the typical amount of angular momentum (in units of $\hbar$) transferred from the laser pulse to the molecule \cite{Fleischer2009}. By amplifying a sequence of 24 pulses in an MPA, we were able to reach kick strengths of up to $P=3$ per pulse ($2 \times 10^{13}$ W/cm$^2$).

%====================================================================
%                    RESULTS
%====================================================================

\begin{figure}
\centering
 \includegraphics[width=1.0\columnwidth]{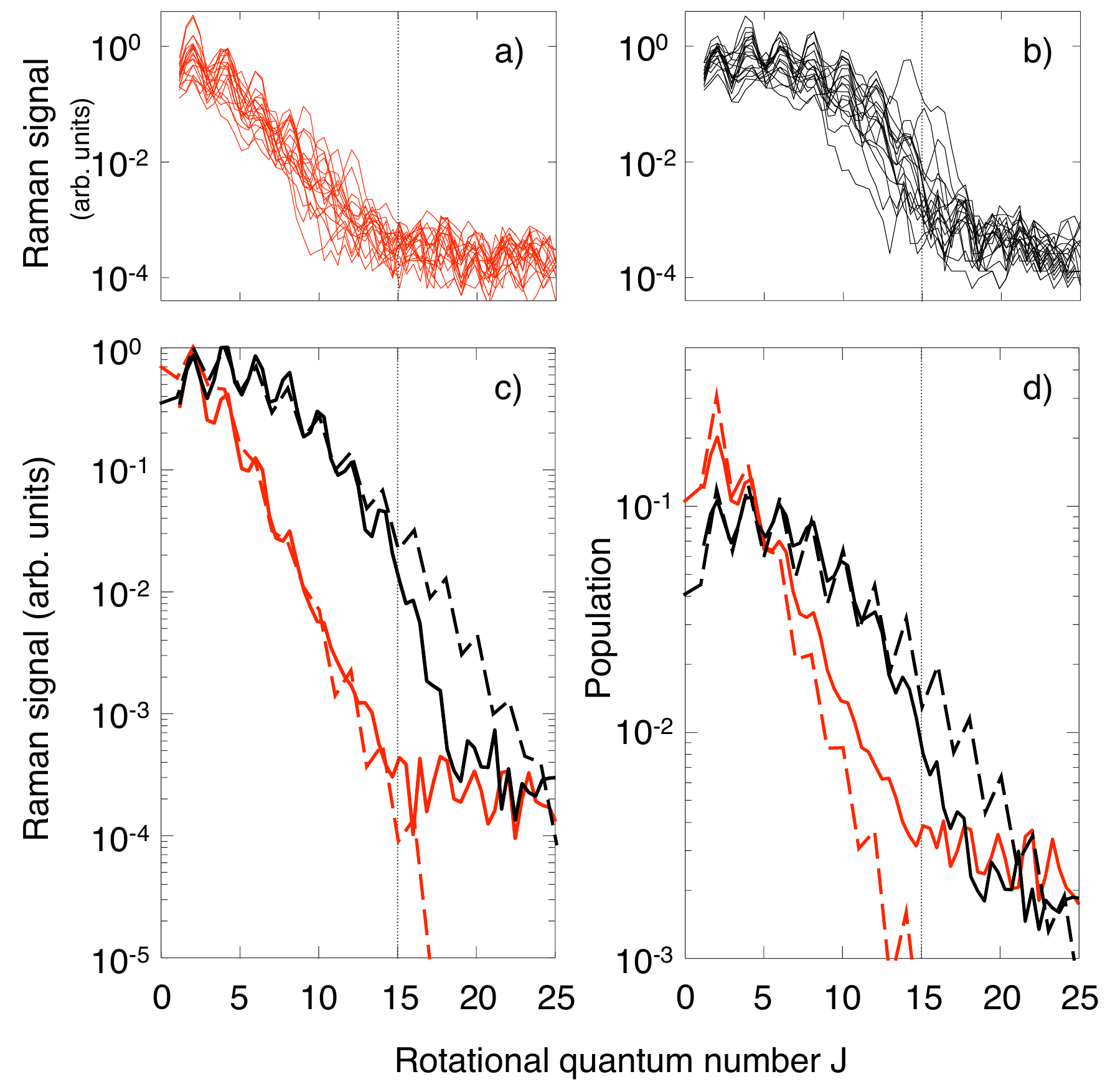}
     \caption{ (color online) Rotational Raman spectrum of nitrogen molecules excited with a train of $N=24$ pulses at a kick strength of $P=2.3$ for 20 different periodic (\textbf{a}) and non-periodic (\textbf{b}) sequences.
    (\textbf{c}) The average experimental distributions (solid lines) are compared to the numerical simulations (dashed lines) for both the periodic (lower red lines)
and the non-periodic (upper black lines) pulse trains. The dotted vertical line represents the excitation limit due to the finite pulse duration.
     (\textbf{d}) The exact calculated populations (dashed lines) and the approximate populations (solid lines), retrieved from the experimental Raman signal as discussed in the text. }
  \vskip -.1truein
  \label{Fig:AL}
\end{figure}

% ------- Results 1:  AL vs Diffusive growth-------------------------

Figure~\ref{Fig:AL}(\textbf{a}) shows a set of 20 Raman spectra, obtained with 20 different \textit{periodic} pulse trains. Localized states in the quantum kicked rotor are known to exist away from the quantum resonances, i.e. when the time between kicks is not equal to a rational fraction of the revival period $T_\text{rev}=(2Bc)^{-1}$ \cite{Fishman1982}. To satisfy this condition, we chose 10 evenly spaced pulse train periods $T$ in each of the two intervals, $10/13<T/T_\text{rev}<5/6$ and $7/8<T/T_\text{rev}<13/14$, with $T_\text{rev}=8.38$~ps for molecular nitrogen $^{14}$N$_2$. The Raman frequency shift (horizontal axis) has been converted to the rotational quantum number $J$.

All of the observed Raman signals $I_J$ decay exponentially across 4 orders of magnitude and 15 rotational states, independent of the train period. The average Raman signal is plotted with the solid red line in Fig.~\ref{Fig:AL}(\textbf{c}). The remaining oscillations are a consequence of the nuclear spin statistics of nitrogen, which dictates the 2:1 ratio for the two independent rotational progressions consisting of only even and only odd values of angular momentum. In Fig.~\ref{Fig:AL}(\textbf{d}), the solid red line illustrates the distribution of the rotational population, extracted from the average Raman signal according to $P_J \propto \sqrt{I_J}$. An evident exponential shape is a hallmark of Anderson localization in this true QKR system.

To confirm the coherent nature of the observed localization, we repeat the same measurement with a set of 20 \textit{non-periodic} pulse trains. The kick strength is set to the same value of $P=2.3$ per pulse, but the time intervals between the 24~pulses in each train is randomly distributed around the mean value of $0.85 T_\text{rev}$ with a standard deviation of $33\%$. Here, all the individual Raman spectra, their average and the population distribution retrieved from it (solid black lines in Fig.~\ref{Fig:AL}(\textbf{b}), (\textbf{c}) and (\textbf{d}), respectively) show a qualitatively different non-exponential shape. As expected for a quantum kicked rotor, Anderson localization is destroyed by the timing noise and the classical diffusion, with its characteristic Gaussian distribution of angular momentum, is recovered.

\begin{figure*}
\centering
 \includegraphics[width=1.0\textwidth]{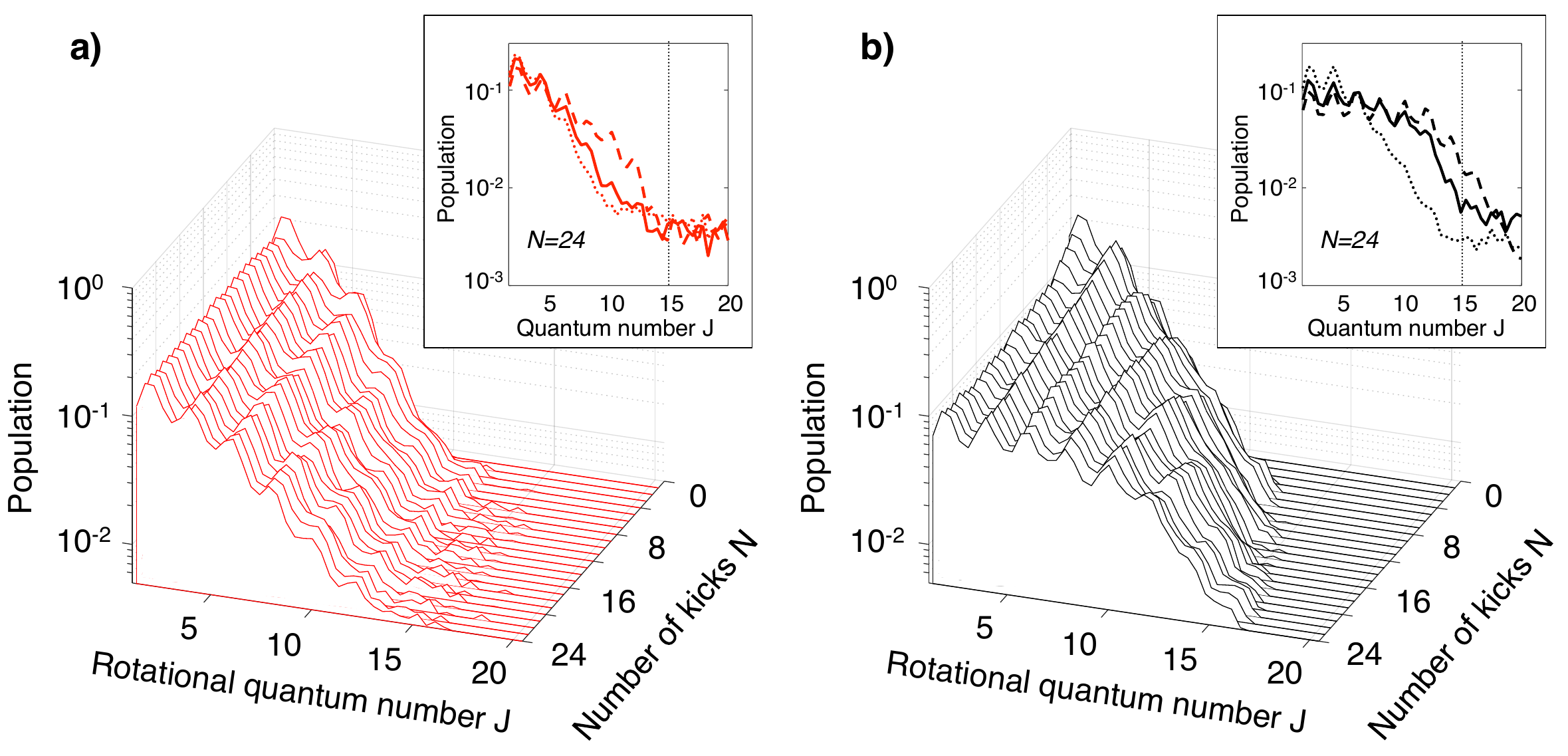}
     \caption{ (color online)
     Evolution of the molecular angular momentum distribution with the number of kicks $N$ (each of strength $P=2.3$) for a periodic (\textbf{a}) and non-periodic (\textbf{b}) excitation. The insets show the corresponding distributions after 24 pulses as a function of the kick strength: $P=1$ (dotted line),  $P=2$ (solid line) and  $P=3$ (dashed line). The dotted vertical line represents the excitation limit due to the finite pulse duration. }
  \vskip -.1truein
  \label{Fig:Waterfall}
\end{figure*}

% ------- Results 2:  Comparison to numerical simulations------------

In Figures \ref{Fig:AL}(\textbf{c}) and (\textbf{d}), we also compare our experimental data to the results of numerical simulations, shown with dashed lines. The latter are carried out by solving the Schr{\"o}dinger equation for nitrogen molecules interacting with a sequence of $\delta$-kicks. We calculate the complex amplitudes $c_{J,M}$ of all rotational states in the wave packet created from each initially populated state $|J',M'\rangle$. Averaging over the initial thermal mixture, we simulate the expected Raman signals $I_J \propto \langle \sum_M |c^{*}_{J,M} c^{}_{J+2,M}|^2 \rangle_{_{J',M'}}$, and find the exact populations $P_J=\langle \sum_M |c_{J,M}|^2 \rangle_{_{J',M'}}$.

In the case of a periodic sequence of kicks, the observed Raman line shape [Fig.\ref{Fig:AL}(\textbf{c})] is in good agreement with the numerical result down to the instrumental noise floor around $I_J \approx 10^{-4}$. Calculated populations [Fig.\ref{Fig:AL}(\textbf{d})] demonstrate the anticipated exponential decay with the rotational quantum number, but deviate slightly from the experimentally retrieved distribution. We attribute this discrepancy to the small finite thermal width of the initial rotational distribution, not accounted for in approximating the populations by $\sqrt{I_J}$, as discussed earlier.

When the timing noise is simulated numerically, both the calculated Raman response and the population distributions show a non-exponential shape and match the experimental observations below $J\approx 15$. The disagreement at higher values of angular momentum is because of the final duration of our laser pulses (130~fs FWHM), which is not taken into account in the simulations. At $J \geqslant 15$ (i.e. to the right of the dotted vertical line), a nitrogen molecule rotates by $\gtrsim 90$\textdegree\ during the length of the pulse, which lowers its effective kick strength and suppresses further rotational excitation.

% ------- Results 3: Localization length as a function of N, P ------

Figure~\ref{Fig:Waterfall} shows the evolution of the rotational distribution with the number of kicks $N$. For the case of a periodic pulse train illustrated in Fig.\ref{Fig:Waterfall}(\textbf{a}), the distribution becomes exponential within a few kicks and hardly changes after that. In sharp contrast, the line shape in Fig.\ref{Fig:Waterfall}(\textbf{b}) remains non-exponential and keeps broadening with increasing $N$ in the case of a non-periodically kicked molecule. This behavior demonstrates the destruction of Anderson localization by timing noise and clearly distinguishes it from other mechanisms of suppressed rotational excitation.

The dependence of the rotational distribution on the strength of periodic and non-periodic kicks is shown in the two respective insets of Fig.\ref{Fig:Waterfall}. As expected for a periodically kicked quantum rotor, the localization length grows with increasing $P$, while the line shape remains exponential below the cutoff value of $J\approx 15$ discussed earlier. For each kick strength, the non-exponential distribution after a noisy pulse sequence lies well above its localized counterpart, despite being equally affected by the cutoff, and thus confirming the universality of the observed dynamics.

\begin{figure}
\centering
 \includegraphics[width=1.0\columnwidth]{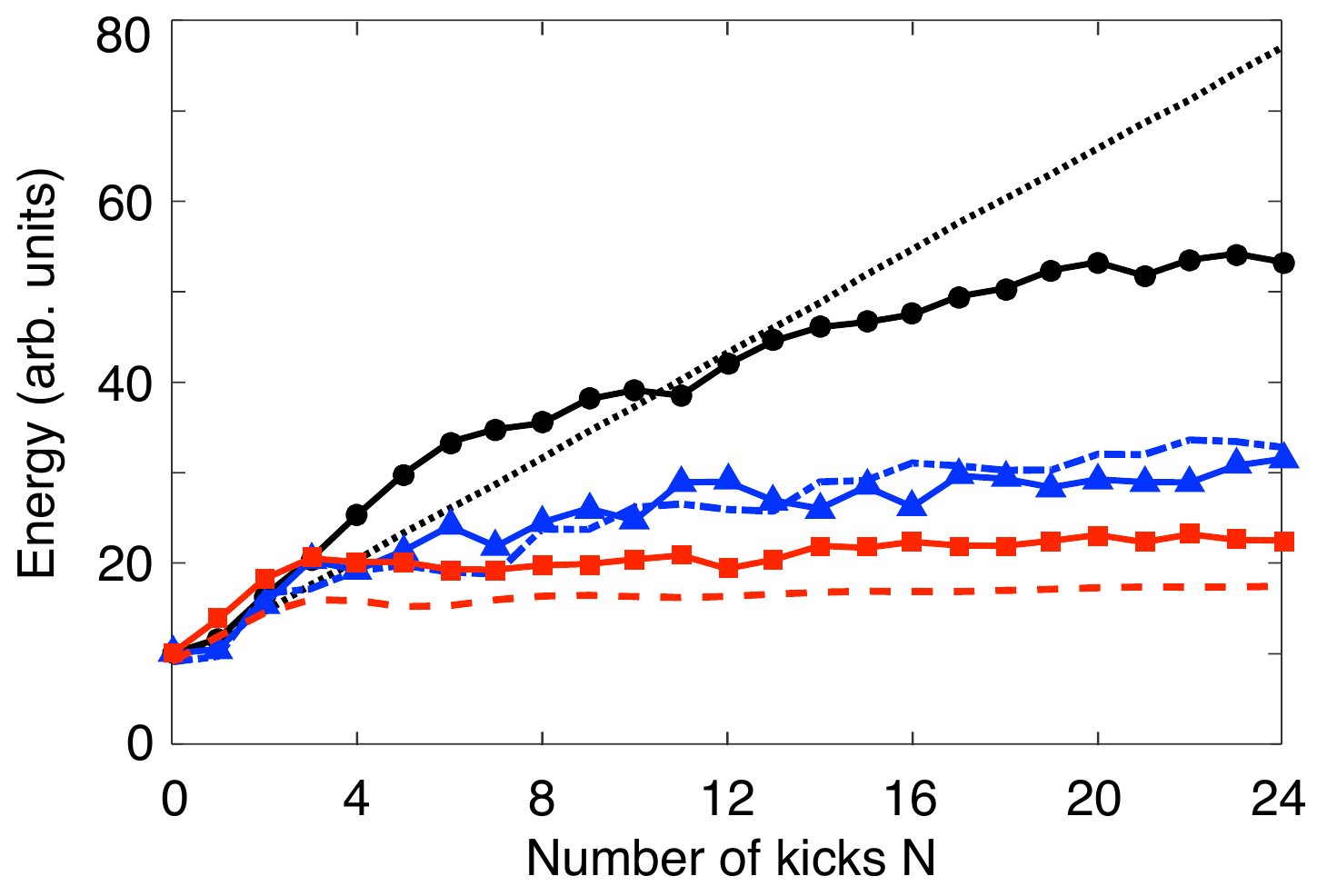}
     \caption{ (color online)
Rotational energy as a function of the number of kicks $N$ with a mean strength of $P=2.3$. Compared are the experimentally retrieved energies (connected symbols) with the numerically calculated ones (lines), for a periodic sequence (red squares, dashed line), and the same sequence after the introduction of amplitude noise (blue triangles, dash-dotted line) or timing noise (black circles, dotted line).}
  \vskip -.1truein
  \label{Fig:Energy}
\end{figure}

% ------- Results 4: Suppressed energy growth -----------------------

Owing to our state resolved detection, the total rotational energy of a molecule can be calculated as $\sum_J E_J P_J$, with populations $P_J$ extracted from the observed Raman spectra $I_J$. The rotational energy is plotted as a function of the number of kicks for multiple excitation scenarios in Fig.\ref{Fig:Energy}. For periodic kicking, the retrieved energy (red squares) increases during the first 3 kicks, after which its further growth is completely suppressed - a prominent feature of Anderson localization in the QKR. Breaking the periodicity of the pulse sequence with timing noise results in the recovery of the classical diffusion, manifested by the continuously increasing rotational energy of the rotors (black circles). The sub-linear growth rate is due to the finite duration of our laser pulses, mentioned earlier. As expected, Anderson localization is also susceptible to amplitude noise. When the amplitudes of a periodic pulse sequence randomly vary with a standard deviation of $41\%$, the steady energy growth is again revived (blue triangles). The numerically calculated rotational energies for the three cases of a periodic, a non-periodic and a noisy-amplitude pulse train are shown with the dashed red, dotted black and dash-dotted blue lines, respectively.

%====================================================================
%                         CONCLUSION
%====================================================================

In summary, we have demonstrated experimentally the effect of Anderson localization in a system of true quantum kicked rotors - a gas of nitrogen molecules exposed to a periodic sequence of intense laser pulses. Cold initial conditions and a high-sensitivity state-resolved detection enabled us to observe the distribution of the molecular angular momenta evolving into an exponential line shape, characteristic of a localized state. The suppressed growth of rotational energy, and the noise-induced recovery of the classical diffusion have also been presented. Our work complements previous studies of the QKR in a system of cold atoms, and opens new opportunities for investigating quantum phenomena which are unique to true rotors, e.g. edge localization \cite{Floss2015b} or the effects of the centrifugal distortion and rotational decoherence on QKR dynamics \cite{Floss2014}. Exploring the possibility of quantum coherent control in these classically chaotic molecular systems is also of great interest \cite{Gong2005}.

%-------- Acknowledgement -------------------------------------------
The authors would like to thank J.~Flo{\ss} and I.~Sh.~Averbukh for stimulating discussions and J.~Flo{\ss} for his help with numerical calculations.

%------------------------Bibliography--------------------------------
%\bibliography{AndersonLocalization}

\begin{thebibliography}{27}
\expandafter\ifx\csname natexlab\endcsname\relax\def\natexlab#1{#1}\fi
\expandafter\ifx\csname bibnamefont\endcsname\relax
  \def\bibnamefont#1{#1}\fi
\expandafter\ifx\csname bibfnamefont\endcsname\relax
  \def\bibfnamefont#1{#1}\fi
\expandafter\ifx\csname citenamefont\endcsname\relax
  \def\citenamefont#1{#1}\fi
\expandafter\ifx\csname url\endcsname\relax
  \def\url#1{\texttt{#1}}\fi
\expandafter\ifx\csname urlprefix\endcsname\relax\def\urlprefix{URL }\fi
\providecommand{\bibinfo}[2]{#2}
\providecommand{\eprint}[2][]{\url{#2}}

\bibitem[{\citenamefont{Casati et~al.}(1979)\citenamefont{Casati, Chirikov,
  Izraelev, and Ford}}]{Casati1979}
\bibinfo{editor}{\bibfnamefont{G.}~\bibnamefont{Casati}},
  \bibinfo{editor}{\bibfnamefont{B.}~\bibnamefont{Chirikov}},
  \bibinfo{editor}{\bibfnamefont{F.}~\bibnamefont{Izraelev}}, \bibnamefont{and}
  \bibinfo{editor}{\bibfnamefont{J.}~\bibnamefont{Ford}}, in
  \emph{\bibinfo{title}{Stochastic Behavior in Classical and Quantum
  Hamiltonian Systems, Lecture Notes in Physics}}, Vol.~\bibinfo{volume}{93}, edited by \bibinfo{editor}{\bibfnamefont{G.}~\bibnamefont{Casati}} \bibnamefont{and}
  \bibinfo{editor}{\bibfnamefont{J.}~\bibnamefont{Ford}}
(\bibinfo{publisher}{Springer}, \bibinfo{address}{Berlin},
  \bibinfo{year}{1979}), pp. \bibinfo{pages}{334--352}.

\bibitem[{\citenamefont{Izrailev and Shepelyanskii}(1980)}]{Izrailev1980}
\bibinfo{author}{\bibfnamefont{F.~M.} \bibnamefont{Izrailev}} \bibnamefont{and}
  \bibinfo{author}{\bibfnamefont{D.~L.} \bibnamefont{Shepelyanskii}},
  \bibinfo{journal}{Theor. Math. Phys.}
  \textbf{\bibinfo{volume}{43}}, \bibinfo{pages}{553} (\bibinfo{year}{1980}).

\bibitem[{\citenamefont{Anderson}(1958)}]{Anderson1958}
\bibinfo{author}{\bibfnamefont{P.~W.} \bibnamefont{Anderson}},
  \bibinfo{journal}{Phys. Rev.} \textbf{\bibinfo{volume}{109}},
  \bibinfo{pages}{1492} (\bibinfo{year}{1958}).
  
\bibitem[{\citenamefont{Fishman et~al.}(1982)\citenamefont{Fishman, Grempel,
  and Prange}}]{Fishman1982}
\bibinfo{author}{\bibfnamefont{S.}~\bibnamefont{Fishman}},
  \bibinfo{author}{\bibfnamefont{D.~R.} \bibnamefont{Grempel}},
  \bibnamefont{and} \bibinfo{author}{\bibfnamefont{R.~E.}
  \bibnamefont{Prange}}, \bibinfo{journal}{Phys. Rev. Lett.}
  \textbf{\bibinfo{volume}{49}}, \bibinfo{pages}{509} (\bibinfo{year}{1982}).

\bibitem[{\citenamefont{Moore et~al.}(1995)\citenamefont{Moore, Robinson,
  Bharucha, Sundaram, and Raizen}}]{Moore1995}
\bibinfo{author}{\bibfnamefont{F.~L.} \bibnamefont{Moore}},
  \bibinfo{author}{\bibfnamefont{J.~C.} \bibnamefont{Robinson}},
  \bibinfo{author}{\bibfnamefont{C.~F.} \bibnamefont{Bharucha}},
  \bibinfo{author}{\bibfnamefont{B.}~\bibnamefont{Sundaram}}, \bibnamefont{and}
  \bibinfo{author}{\bibfnamefont{M.~G.} \bibnamefont{Raizen}},
  \bibinfo{journal}{Phys. Rev. Lett.} \textbf{\bibinfo{volume}{75}},
  \bibinfo{pages}{4598} (\bibinfo{year}{1995}).

\bibitem[{\citenamefont{Raizen}(1999)}]{Raizen1999}
\bibinfo{author}{\bibfnamefont{M.~G.} \bibnamefont{Raizen}},
  \bibinfo{booktitle}{Adv. At. Mol. Opt. Phys.} \textbf{\bibinfo{volume}{41}},
  \bibinfo{pages}{43} (\bibinfo{year}{1999}).

\bibitem[{\citenamefont{Bluemel et~al.}(1986)\citenamefont{Bluemel, Fishman,
  and Smilansky}}]{Bluemel1986}
\bibinfo{author}{\bibfnamefont{R.}~\bibnamefont{Bluemel}},
  \bibinfo{author}{\bibfnamefont{S.}~\bibnamefont{Fishman}}, \bibnamefont{and}
  \bibinfo{author}{\bibfnamefont{U.}~\bibnamefont{Smilansky}},
  \bibinfo{journal}{J. Chem. Phys.}
  \textbf{\bibinfo{volume}{84}}, \bibinfo{pages}{2604} (\bibinfo{year}{1986}).

\bibitem[{\citenamefont{Gong and Brumer}(2001)}]{Gong2001}
\bibinfo{author}{\bibfnamefont{J.}~\bibnamefont{Gong}} \bibnamefont{and}
  \bibinfo{author}{\bibfnamefont{P.}~\bibnamefont{Brumer}},
  \bibinfo{journal}{J. Chem. Phys.}
  \textbf{\bibinfo{volume}{115}}, \bibinfo{pages}{3590} (\bibinfo{year}{2001}).

\bibitem[{\citenamefont{Floss and Averbukh}(2012)}]{Floss2012}
\bibinfo{author}{\bibfnamefont{J.}~\bibnamefont{Flo\ss}} \bibnamefont{and}
  \bibinfo{author}{\bibfnamefont{I.~Sh.} \bibnamefont{Averbukh}},
  \bibinfo{journal}{Phys. Rev. A} \textbf{\bibinfo{volume}{86}},
  \bibinfo{pages}{021401} (\bibinfo{year}{2012}).

\bibitem[{\citenamefont{Matsuoka}(2015)}]{Matsuoka2015}
\bibinfo{author}{\bibfnamefont{L.}~\bibnamefont{Matsuoka}},
  \bibinfo{journal}{Phys. Rev. A} \textbf{\bibinfo{volume}{91}},
  \bibinfo{pages}{043420} (\bibinfo{year}{2015}).

\bibitem[{\citenamefont{Floss et~al.}(2013)\citenamefont{Floss, Fishman, and
  Averbukh}}]{Floss2013}
\bibinfo{author}{\bibfnamefont{J.}~\bibnamefont{Flo\ss}},
  \bibinfo{author}{\bibfnamefont{S.}~\bibnamefont{Fishman}}, \bibnamefont{and}
  \bibinfo{author}{\bibfnamefont{I.~Sh.} \bibnamefont{Averbukh}},
  \bibinfo{journal}{Phys. Rev. A} \textbf{\bibinfo{volume}{88}},
  \bibinfo{pages}{023426} (\bibinfo{year}{2013}).

\bibitem[{\citenamefont{Floss and Averbukh}(2014)}]{Floss2014}
\bibinfo{author}{\bibfnamefont{J.}~\bibnamefont{Flo\ss}} \bibnamefont{and}
  \bibinfo{author}{\bibfnamefont{I.~Sh.} \bibnamefont{Averbukh}},
  \bibinfo{journal}{Phys. Rev. Lett.} \textbf{\bibinfo{volume}{113}},
  \bibinfo{pages}{043002} (\bibinfo{year}{2014}).

\bibitem[{\citenamefont{Floss and Averbukh}(2015)}]{Floss2015b}
\bibinfo{author}{\bibfnamefont{J.}~\bibnamefont{Flo\ss}} \bibnamefont{and}
  \bibinfo{author}{\bibfnamefont{I.~Sh.} \bibnamefont{Averbukh}},
  \bibinfo{journal}{Phys. Rev. E} \textbf{\bibinfo{volume}{91}},
  \bibinfo{pages}{052911} (\bibinfo{year}{2015}).

\bibitem[{\citenamefont{Cryan et~al.}(2009)\citenamefont{Cryan, Bucksbaum, and
  Coffee}}]{Cryan2009}
\bibinfo{author}{\bibfnamefont{J.~P.} \bibnamefont{Cryan}},
  \bibinfo{author}{\bibfnamefont{P.~H.} \bibnamefont{Bucksbaum}},
  \bibnamefont{and} \bibinfo{author}{\bibfnamefont{R.~N.}
  \bibnamefont{Coffee}}, \bibinfo{journal}{Phys. Rev. A}
  \textbf{\bibinfo{volume}{80}}, \bibinfo{pages}{063412}
  (\bibinfo{year}{2009}).

\bibitem[{\citenamefont{Zhdanovich et~al.}(2012)\citenamefont{Zhdanovich,
  Bloomquist, FloÃŸ, Averbukh, Hepburn, and Milner}}]{Zhdanovich2012}
\bibinfo{author}{\bibfnamefont{S.}~\bibnamefont{Zhdanovich}},
  \bibinfo{author}{\bibfnamefont{C.}~\bibnamefont{Bloomquist}},
  \bibinfo{author}{\bibfnamefont{J.}~\bibnamefont{Flo\ss}},
  \bibinfo{author}{\bibfnamefont{I.~Sh.} \bibnamefont{Averbukh}},
  \bibinfo{author}{\bibfnamefont{J.~W.} \bibnamefont{Hepburn}},
  \bibnamefont{and} \bibinfo{author}{\bibfnamefont{V.}~\bibnamefont{Milner}},
  \bibinfo{journal}{Phys. Rev. Lett.} \textbf{\bibinfo{volume}{109}},
  \bibinfo{pages}{043003} (\bibinfo{year}{2012}).

\bibitem[{\citenamefont{Floss et~al.}(2015)\citenamefont{Floss, Kamalov,
  Averbukh, and Bucksbaum}}]{Floss2015a}
\bibinfo{author}{\bibfnamefont{J.}~\bibnamefont{Flo\ss}},
  \bibinfo{author}{\bibfnamefont{A.}~\bibnamefont{Kamalov}},
  \bibinfo{author}{\bibfnamefont{I.~Sh.} \bibnamefont{Averbukh}},
  \bibnamefont{and} \bibinfo{author}{\bibfnamefont{P.~H.}
  \bibnamefont{Bucksbaum}}, \bibinfo{journal}{Phys. Rev. Lett.}
  \textbf{\bibinfo{volume}{115}}, \bibinfo{pages}{203002}
  (\bibinfo{year}{2015}).

\bibitem[{\citenamefont{Kamalov et~al.}(2015)\citenamefont{Kamalov, Broege, and
  Bucksbaum}}]{Kamalov2015}
\bibinfo{author}{\bibfnamefont{A.}~\bibnamefont{Kamalov}},
  \bibinfo{author}{\bibfnamefont{D.~W.} \bibnamefont{Broege}},
  \bibnamefont{and} \bibinfo{author}{\bibfnamefont{P.~H.}
  \bibnamefont{Bucksbaum}}, \bibinfo{journal}{Phys. Rev. A}
  \textbf{\bibinfo{volume}{92}}, \bibinfo{pages}{013409}
  (\bibinfo{year}{2015}).

\bibitem[{\citenamefont{Klappauf et~al.}(1998)\citenamefont{Klappauf, Oskay,
  Steck, and Raizen}}]{Klappauf1998}
\bibinfo{author}{\bibfnamefont{B.~G.} \bibnamefont{Klappauf}},
  \bibinfo{author}{\bibfnamefont{W.~H.} \bibnamefont{Oskay}},
  \bibinfo{author}{\bibfnamefont{D.~A.} \bibnamefont{Steck}}, \bibnamefont{and}
  \bibinfo{author}{\bibfnamefont{M.~G.} \bibnamefont{Raizen}},
  \bibinfo{journal}{Phys. Rev. Lett.} \textbf{\bibinfo{volume}{81}},
  \bibinfo{pages}{1203} (\bibinfo{year}{1998}).

\bibitem[{\citenamefont{Ammann et~al.}(1998)\citenamefont{Ammann, Gray,
  Shvarchuck, and Christensen}}]{Ammann1998}
\bibinfo{author}{\bibfnamefont{H.}~\bibnamefont{Ammann}},
  \bibinfo{author}{\bibfnamefont{R.}~\bibnamefont{Gray}},
  \bibinfo{author}{\bibfnamefont{I.}~\bibnamefont{Shvarchuck}},
  \bibnamefont{and}
  \bibinfo{author}{\bibfnamefont{N.}~\bibnamefont{Christensen}},
  \bibinfo{journal}{Phys. Rev. Lett.} \textbf{\bibinfo{volume}{80}},
  \bibinfo{pages}{4111} (\bibinfo{year}{1998}).

\bibitem[{\citenamefont{Milner et~al.}(2000)\citenamefont{Milner, Steck, Oskay,
  and Raizen}}]{Milner2000}
\bibinfo{author}{\bibfnamefont{V.}~\bibnamefont{Milner}},
  \bibinfo{author}{\bibfnamefont{D.~A.} \bibnamefont{Steck}},
  \bibinfo{author}{\bibfnamefont{W.~H.} \bibnamefont{Oskay}}, \bibnamefont{and}
  \bibinfo{author}{\bibfnamefont{M.~G.} \bibnamefont{Raizen}},
  \bibinfo{journal}{Phys. Rev. E} \textbf{\bibinfo{volume}{61}},
  \bibinfo{pages}{7223} (\bibinfo{year}{2000}).

\bibitem[{\citenamefont{Oskay et~al.}(2003)\citenamefont{Oskay, Steck, and
  Raizen}}]{Oskay2003}
\bibinfo{author}{\bibfnamefont{W.~H.} \bibnamefont{Oskay}},
  \bibinfo{author}{\bibfnamefont{D.~A.} \bibnamefont{Steck}}, \bibnamefont{and}
  \bibinfo{author}{\bibfnamefont{M.~G.} \bibnamefont{Raizen}},
  \bibinfo{journal}{Chaos, Solitons \& Fractals} \textbf{\bibinfo{volume}{16}},
  \bibinfo{pages}{409} (\bibinfo{year}{2003}).

\bibitem[{\citenamefont{Bitter and Milner}(2016{\natexlab{a}})}]{Bitter2016a}
\bibinfo{author}{\bibfnamefont{M.}~\bibnamefont{Bitter}} \bibnamefont{and}
  \bibinfo{author}{\bibfnamefont{V.}~\bibnamefont{Milner}},
  \bibinfo{journal}{Appl. Opt.} \textbf{\bibinfo{volume}{55}},
  \bibinfo{pages}{830} (\bibinfo{year}{2016}{\natexlab{a}}).

\bibitem[{\citenamefont{Weiner}(2000)}]{Weiner2000}
\bibinfo{author}{\bibfnamefont{A.~M.} \bibnamefont{Weiner}},
  \bibinfo{journal}{Rev. Sci. Instrum.}
  \textbf{\bibinfo{volume}{71}}, \bibinfo{pages}{1929} (\bibinfo{year}{2000}).

\bibitem{Temperature}
The temperature was retrieved numerically by fitting the rotational spectrum obtained after a single weak pulse.


\bibitem[{\citenamefont{Korech et~al.}(2013)\citenamefont{Korech, Steinitz,
  Gordon, Averbukh, and Prior}}]{Korech2013}
\bibinfo{author}{\bibfnamefont{O.}~\bibnamefont{Korech}},
  \bibinfo{author}{\bibfnamefont{U.}~\bibnamefont{Steinitz}},
  \bibinfo{author}{\bibfnamefont{R.~J.} \bibnamefont{Gordon}},
  \bibinfo{author}{\bibfnamefont{I.~Sh.} \bibnamefont{Averbukh}},
  \bibnamefont{and} \bibinfo{author}{\bibfnamefont{Y.}~\bibnamefont{Prior}},
  \bibinfo{journal}{Nat. Photon.} \textbf{\bibinfo{volume}{7}},
  \bibinfo{pages}{711} (\bibinfo{year}{2013}).

\bibitem[{\citenamefont{Korobenko et~al.}(2014)\citenamefont{Korobenko, Milner,
  and Milner}}]{Korobenko2014a}
\bibinfo{author}{\bibfnamefont{A.}~\bibnamefont{Korobenko}},
  \bibinfo{author}{\bibfnamefont{A.~A.} \bibnamefont{Milner}},
  \bibnamefont{and} \bibinfo{author}{\bibfnamefont{V.}~\bibnamefont{Milner}},
  \bibinfo{journal}{Phys. Rev. Lett.} \textbf{\bibinfo{volume}{112}},
  \bibinfo{pages}{113004} (\bibinfo{year}{2014}).


\bibitem[{\citenamefont{Cohen}(1991)}]{Cohen1991}
\bibinfo{author}{\bibfnamefont{D.}~\bibnamefont{Cohen}},
  \bibinfo{journal}{Phys. Rev. A} \textbf{\bibinfo{volume}{44}},
  \bibinfo{pages}{2292} (\bibinfo{year}{1991}).

\bibitem{approximation}
We have verified this approximation numerically.
  

\bibitem[{\citenamefont{Bitter and Milner}(2016{\natexlab{b}})}]{Bitter2016b}
\bibinfo{author}{\bibfnamefont{M.}~\bibnamefont{Bitter}} \bibnamefont{and}
  \bibinfo{author}{\bibfnamefont{V.}~\bibnamefont{Milner}},
  \bibinfo{journal}{Phys. Rev. A} \textbf{\bibinfo{volume}{93}},
  \bibinfo{pages}{013420} (\bibinfo{year}{2016}{\natexlab{b}}).

\bibitem{Pcalibration}
The details of this method will be discussed in a future publication.


\bibitem[{\citenamefont{Fleischer et~al.}(2009)\citenamefont{Fleischer,
  Khodorkovsky, Prior, and Averbukh}}]{Fleischer2009}
\bibinfo{author}{\bibfnamefont{S.}~\bibnamefont{Fleischer}},
  \bibinfo{author}{\bibfnamefont{Y.}~\bibnamefont{Khodorkovsky}},
  \bibinfo{author}{\bibfnamefont{Y.}~\bibnamefont{Prior}}, \bibnamefont{and}
  \bibinfo{author}{\bibfnamefont{I.~Sh.} \bibnamefont{Averbukh}},
  \bibinfo{journal}{New J. Phys.} \textbf{\bibinfo{volume}{11}},
  \bibinfo{pages}{105039} (\bibinfo{year}{2009}).

\bibitem[{\citenamefont{Gong and Brumer}(2005)}]{Gong2005}
\bibinfo{author}{\bibfnamefont{J.}~\bibnamefont{Gong}} \bibnamefont{and}
  \bibinfo{author}{\bibfnamefont{P.}~\bibnamefont{Brumer}},
  \bibinfo{journal}{Annu. Rev. Phys. Chem.}
  \textbf{\bibinfo{volume}{56}}, \bibinfo{pages}{1} (\bibinfo{year}{2005}).

\end{thebibliography}
%\end{document}

\end{document}